# Highly sensitive silver decorated-graphene oxide-silicon nanowires hybrid SERS sensors for trace level detection of environmental pollutants

Kais Daoudi[1, 2, 3*], Mounir Gaidi[1, 2, 4**], Soumya Columbus[2,5], Mohammed Shameer[2] and Hussain Alawadhi[1, 2]

[1)] Department of Applied Physics and Astronomy, University of Sharjah, P. O. Box 27272 Sharjah, United Arab Emirates

[2)] Centre for Advanced Materials Research, Research Institute of Sciences and Engineering, University of Sharjah, P. O. Box 27272 Sharjah, United Arab Emirates

[3)] Laboratory of Nanomaterials Nanotechnology and Energy (2NE), Faculty of Sciences of Tunis, University of Tunis El Manar, 2092, Tunis, Tunisia

[4)] Laboratoire de Photovoltaïque, Centre de Recherches et des Technologies de l'Energie, Technopole de Borj-Cédria, 2050 Hammam-Lif, Tunisia

[5)] Sharjah Research Academy, P.O. Box 60999, Sharjah, University City, Sharjah, United Arab Emirates

**Abstract**

In this study, we evaluated the sensing performance of silver nanoprism/graphene oxide/silicon nanowires (AgNPr/GO/SiNWs) nanohybrid system for effective detection of organic dye and herbicide residues in freshwater via surface-enhanced Raman scattering (SERS). Homogenous and vertically aligned SiNWs have been successfully synthesized using metal-assisted chemical etching technique by varying the etching time from 10 to 30 min. AgNPr/GO/SiNW hybrids were assembled by spin-coating of GO followed by drop-casting deposition of AgNPr. The microstructures of AgNPr/GO/SiNWs are strongly affected by the etching time of SiNWs, which in turn affected its SERS performance when rhodamine 6G is used as an analyte molecule. Owing to the synergetic effects of GO and AgNPr, SERS response of AgNPr/GO/SiNWs composites were found to be superior compared to AgNPr/SiNWs. High efficiency of $6.1 \times 10^{10}$ has been achieved with AgNPr/GO/SiNWs based sensor fabricated with 30-min etched SiNWs for rhodamine 6G. An effective way of rapid detection of drinking water pollutants such as methylene blue and atrazine up to picomolar ($10^{-12}$ M) concentrations was demonstrated using these low-cost AgNPr/GO/SiNWs nanohybrids SERS sensors.

Electronic mail: *kdaoudi@sharjah.ac.ae

\**mkaidi@sharjah.ac.ae

# 1. Introduction

The development of novel strategies to face environmental issues is greatly necessitated in recent years owing to the high susceptibility of natural resources towards pollution. Herbicides and organic dyes are reported to be among the major contaminants that drastically polluted water resources (Robinson et al., 2001; Rämö et al., 2018). Former are extensively employed worldwide in the agricultural field were only very low portions is being directed to the crops during its application. The remaining herbicide residues can diffuse into the soil and reported to contaminate the underground water (Thurman et al., 1992; Arias-Estévez et al., 2008). Atrazine is a typical herbicide, which is characterized by poor biodegradation rate and low water dissolution. It can adversely affect human health by its inherent carcinogenetic and endocrine disruptor effects (Gasnier et al., 2009). Organic dyes constitute an important class of synthetic organic compounds used in many industries, especially textiles. Consequently, they have become common industrial environmental pollutants during their synthesis as well as during the coloring stage. The large-scale production and extensive application of synthetic dyes are stable and toxic; which can cause deleterious effects in ecosystem (Li et al., 2018). Methylene blue (MB) is a common dye particularly used in textile industries, whose presence has been detected in aquatic organisms like fishes (Li et al., 2016a). Being potentially carcinogenic and hazardous to health, it may be detrimental once enter into humans through the food chain. Both atrazine and methylene blue residues were found to interfere with the photosynthesis process, thereby negatively affecting the aquatic ecosystem. As conventional detection techniques including chromatographic methods could not meet the current needs, the development of rapid and efficient monitoring analytical methods attracts a lot of research interests.

Surface-Enhanced Raman Scattering (SERS) is emerging as a promising and powerful tool for detecting chemical or biological molecules with high sensitivity and excellent efficiency. Non-destructively, substances can be detected with ~$10^{14}$ enhancement in Raman signal by adsorbing on metal nanoparticle matrix [8]. To explain the SERS effect, two commonly admitted mechanisms are generally reported, the electromagnetic mechanism and the chemical one. The electromagnetic mechanism depends on the generation of 'hotspots' where Raman signal of target materials could be enhanced significantly due to a strong electromagnetic field creation that results in a significant increase in the cross-section of the Raman scattering [9]. The contribution to the EM enhancement is mainly due to the Metal nanoparticles (MNPs) surface plasmons excited by the incident light. The enhancement is proportional to $|E|^4$ and can get to $10^8$ or more, where $E$ is the intensity of the electromagnetic field (Kneipp et al., 1997). On the other hand, the chemical mechanism is based on a charge transfer between the targeted molecule and the substrate. Because of charge transfer, the positive and negative charge in the molecule become more separated, which means the polarizability of the molecule increases leading to an enhancement in the cross-section of the Raman scattering. The enhancement due to the chemical mechanism is in the range $10 \sim 10^2$. The chemical mechanism is a molecular scale effect, which means it needs the molecule to be in contact with the substrate or very close to it, so that charge transfer can easily occur between the molecule and the substrate (2D materials such as graphene oxide).

Metal nanoparticles (MNPs) such as gold, silver, and copper are widely employed as SERS substrates owing to their localized surface plasmonic (LSP) resonance which gives rise to hot spots [10-12]. Silver nanoparticles are more often employed owing to their low cost and a strong enhancement factor which could be due to large dielectric constant and thermal stability [13]. Among two different strategies of SERS substrates such as (i) colloidal suspension and (ii) immobilized plasmonic nanoparticles in solid support, the later one has more advantages. The continuous movement of NPs within the colloidal suspension can give rise to aggregation, the net SERS effect could get compromised (Panneerselvam et al., 2018). Meanwhile, immobilized NPs on the solid phase found to provide much reliable and reproducible SERS substrates for practical applications. Despite significant research efforts so far, the reproducibility and stability of the SERS substrates remain a major challenge that needs to be addressed [15].

Silicon nanowires (SiNWs) are one-dimensional nanostructures that have known immense research potential in various fields due to their unique electronic, optical, and mechanical properties [16–18]. The huge surface area and nanoscale dimensions made them excellent candidates for sensor applications. Also, SiNWs structures can be fabricated by a simple and cost-effective method like metal-assisted chemical etching (MaCE) process [19,20]. Nanostructure of SiNWs could be modulated by varying experimental parameters like etching time, etching solution concentration, etc. Recent studies have demonstrated that silicon nanowire functionalized with metal nanoparticles can exhibit ultrahigh sensitivity along with excellent reproducibility (Galopin et al., 2009a; He et al., 2011; Cui et al., 2017). These composite structures could show a high enhancement factor, as the hotspots are located at the interface between the metal nanoparticle and the semiconducting silicon nanowires, which could further enhance their detection limit (Shao et al., 2008; Convertino et al., 2016). It has also been shown that localized electromagnetic fields are most intense near regions with large radius of curvatures (e.g SiNWs sharp tips/edges) or having high roughness (Galopin et al., 2009a; Liu et al., 2015). SiNWs nanostructure contributes also efficiently to the light confinement.

Moreover, graphene or graphene oxide (GO)-metal nanoparticle composites were found to have better molecular sensing compared to bare plasmonic nanoparticles (Wang et al., 2013). Such 2D class of materials is atomically flat and empowers effective charge-transfer due to the short separation distance between graphene and probe molecules. As well, the presence of a GO layer into some SERS substrates has proved to generate strong chemical enhancement (Li et al., 2017). Chemical sensing quality of the graphene-based structure is mainly related to its excellent physical, chemical, and electrical properties including large effective surface area, high electron transfer capability, and good chemical stability (Yavari and Koratkar, 2012; Thi et al., 2020). GO can also be coupled with SiNWs to fabricate hybrid SERS substrates with superior properties (Dimiev and Tour, 2014; Li et al., 2016b; Nien et al., 2016).

Despite the great potential of GO in sensing applications, few studies have explored the feasibility of graphene oxide-based composite SERS substrates. Herein, we designed silver nanoprisms/graphene oxide/silicon nanowires (AgNPr/GO/SiNWs) nanohybrid as SERS based chemical sensors for monitoring of environmental pollutants at trace levels. The microstructure of the nanocomposite substrates has been extensively investigated in terms of etching time and correlated to their SERS responses. Our results point out that 30-min etched SiNWs decorated with GO and AgNPr provide the optimum microstructure for getting the highest SERS response. A maximum efficiency of 6.1 x $10^{10}$ has been achieved. It is worth mentioning that the detection of environmental hazards like organic dyes and herbicides with extremely high sensitivity levels was demonstrated with AgNPr/GO/SiNWs composites.

## 2. Materials and methods

### 2.1. Materials

Silicon wafers (p-type, 100) having thicknesses of 675 ± 25 µm were received from TED PELLA, USA. Hydrofluoric acid (ACS grade, 48%) hydrogen peroxide ($H_2O_2$, 35 %), and anhydrous glacial acetic acid (ACS grade, 100%) were obtained from Merck, Germany. Nitric acid (ISO grade, 65%), Rhodamine 6G (R6G, R & D grade), and atrazine (R & D grade) from Sigma-Aldrich, MB (> 98 % purity) was also obtained from IBI scientific, IA were also utilized for the study.

### 2.2. Preparation of SiNWs

SiNWs samples were synthesized by metal-assisted chemical etching (MACE) method using single side polished silicon wafers (1.5 cm x 1.5 cm). Firstly, Si wafers were ultrasonically cleaned in acetone and ethanol for 5 min each, followed by dipping in an acid solution of $HNO_3$, HF and $CH_3COOH$ (64%, 16%, and 20% respectively) for 1 min. Cleaned wafers were thoroughly rinsed with deionized water and dried. Subsequently, substrates were immersed in an aqueous solution of HF (4.8 M) and $AgNO_3$ (0.02 M) for 1 minute to induce the deposition of silver nanoparticles on it. In the etching step, the substrates were immersed in an aqueous solution of $HF/H_2O_2/H_2O$ (4/1/20 volume ratio) for 10, 20, and 30 min after which they turn into a brownish color. The samples were then immersed in $HNO_3/H_2O$ solution (1:1) for 10 min to remove silver particles. After final rinsing and drying operations, the substrates turned black with a layer of SiNWs.

### 2.3. Fabrication of AgNPr/GO/SiNWs SERS sensors

Silver nanoprisms (AgNPr) and Graphene oxide (GO) were respectively synthesized using the chemical reduction process and Hummers' method (Hummers and Offeman, 1958; Zhang et al., 2019). For fabricating AgNPr/GO/SiNWs sensor device, SiNWs was subjected to spin coating by a GO/water solution at 2000 rpm for 30 sec (please mention the GO concentration) followed by drop-casting of AgNPr solution and allowed to dry at 50°C for 10 min. The process is repeated for several times to ensure proper AgNPr density on final AgNPr/GO/SiNWs substrates. For comparison, AgNPr/SiNWs were also fabricated using drop-casting of AgNPr solution.

## 2.4. Methodology

Surface microstructures of SiNWs substrates were studied using field emission scanning electron microscope (FESEM, Thermo scientic Apreo C). Morphology was also analyzed by Atomic force microscopy (AFM, Nanosurf 3000). WxSM software was used for processing AFM images and analyzing data (Horcas et al., 2007). SERS measurements were performed on Raman microscope (Renishaw) using 488 nm laser excitation of power of 100 µW with 50x objective (~1 µm$^2$ spot size) for an integration time of 10 seconds. Raman spectra were recorded for R6G deposited on different composite substrates by keeping laser power as 0.1% and exposure time as 10 seconds. For this, R6G/ethanol solution (7 µl, 5 x 10$^{-6}$ M) was drop casted on the hybrid SiNWs based sensor, and Raman analysis was carried out after ethanol evaporation. For sensitivity and low detection limit exploration, R6G/ethanol (5 x 10$^{-7}$ to 5 x 10$^{-12}$ M), MB/deionized water, and atrazine/deionized water solutions (2 x 10$^{-7}$ to 2 x 10$^{-12}$ M) were prepared, deposited on substrates and Raman spectra were recorded. For atrazine solutions, 20-min ultra-sonication was applied to ensure proper solubility in water.

## 3. Results and Discussion
### 3.1. Effect of etching time on the microstructure of AgNPr/GO/SiNWs nanohydrides

Silicon nanowires were fabricated using MACE process by altering the etching time from 10 to 30 min. FESEM images of 10-min etched SiNWs demonstrated the formation of homogenously distributed, dense, and vertically nanowires aligned (Figure 1a & b). After decorating with silver nanoprisms (AgNPr), The morphologies of the corresponding AgNPr/SiNW composites were also analyzed (Figure 1c & d). The distribution of AgNPr on the bundle top is clearly shown in magnified images (Figure 1c- inset). Furthermore, the topology of AgNPr/GO modified NWs (AgNPr/GO/SiNWs) were also elucidated as shown in Figure 1*e & f*. It can be seen that the nanowire bundles were almost covered with GO layers on which AgNPr clusters were well dispersed. High density of AgNPrs was observed on AgNPr/GO/SiNWs compared to AgNPr/SiNWs composites. This could be due to the presence of GO layers on NWs, which can provide a uniform platform for AgNPr coating while restricting the migration of nanoparticles to the bottom of NWs.

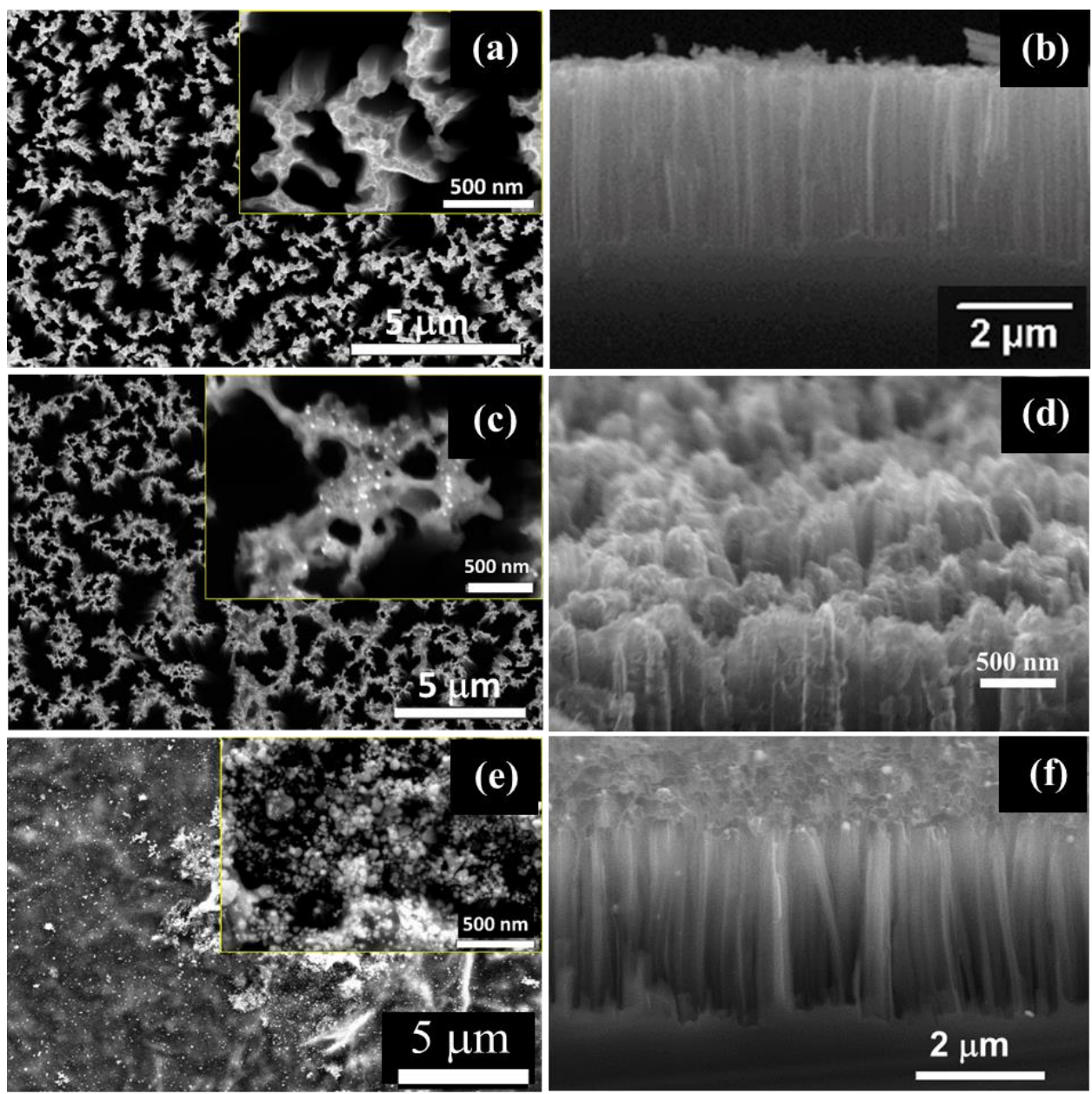

**Figure 1:** Top and cross section FESEM images of (a) & (b) 10-min etched SiNWs (c) & (d) AgNPr/SiNWs and (e) and (f) AgNPr/GO/SiNWs; the corresponding higher magnification images are given as inset.

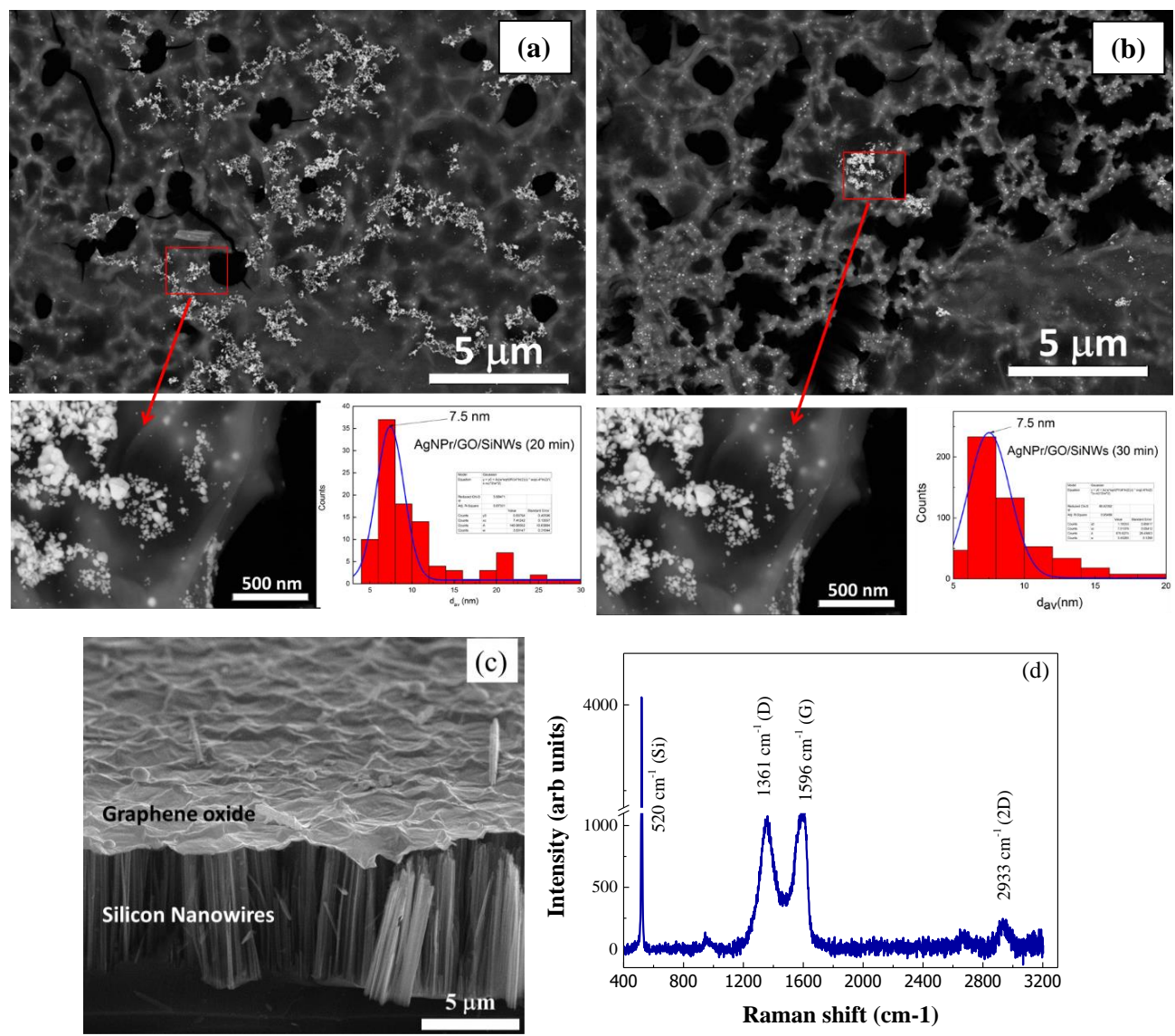

**Figure 2:** FESEM top view of AgNPr/GO/SiNWs structures obtained with an etching time of (a) 20 min (a) and 30 min (b). The corresponding higher magnified FESEM images showing the distributions of the AgNPr with plots of size distributions. (c) and (d) show typical SEM image and Raman spectrum of the GO deposited on SiNWs etched for 20 minutes.

To study the effect of etching time on the AgNPr/GO/SiNWs microstructure of composite, FESEM images of composites prepared at higher etching times were also analyzed (Figure 2). It could be seen that finely oriented, transparent GO layers were present on 20- and 30- min etched NWs. This could be correlated to the corresponding change in the morphology of NWs as a function of etching time. Our earlier studies have demonstrated that an increase in etching time can subsequently enhance the nanowire length (Daoudi et al., 2020). We have found that the NWs length could be tuned from 3.5 to 10.5 µm by altering etching time from 10 to 30 min. The subsequent enhancement in NW length was found to influence the top bundle morphology appreciably since the effective increase in nanowire length can facilitates the Van der Waals forces of attraction within the NWs. This leads to the assembly of more nanowires together at the top, leading to an increase in NW bundle size (Srivastava et al., 2014; Daoudi et al., 2020). The larger bundles found to enhance the better orientation of GO layers within it. Further incorporation of AgNPr on these GO layers facilitates the formation of closely located plasmonic nanoparticle clusters. Analysis of the size distribution showed that the average size of the AgNPr is independent of the etching time. A typical SEM image of the GO freshly deposited on SiNWs etched for 20 minutes is shown in Fig. 2 (c). This result confirms that the orientation of the deposited GO layer accommodates with the surface morphology of the SiNWs substrate. The typical Raman spectra of AgNPr/GO/SiNW composites further confirms the peaks corresponding to Si at 521 cm$^{-1}$ and the D, G and 2D peaks of GO (Fig. 2 (d)).

## 3.2. Effect of etching time and GO modification on SERS sensor performance

Figure 3 shows the effect of etching time on SERS properties prior (a) and after GO incorporation (b). In contrast to the weak Raman signal of R6G on bare Si, AgNPr/Si and AgNPr/SiNWs composites showed fingerprint peaks of R6G at 612, 774, 1362, 1509, and 1650 cm$^{-1}$ (Lai et al., 2018). The plasmonic properties of AgNPr on Si or SiNWs substrates with tuned SERS effect have been thus demonstrated as a function of its microstructure. It could be seen that the enhancement in Raman peaks is much pronounced for AgNPr/SiNWs than AgNPr/Si substrates. The huge surface area of SiNWs enables to accommodate more AgNPr and R6G molecules on it, thereby enhancing the net SERS effect on SiNWs composites (Galopin et al., 2009b). One could clearly observe the trend in varying Raman scattering of R6G on AgNPr/SiNW composites with an increase in etching time (Figure 3a). 10-min etched NWs showed the highest Raman enhancement, which gets subsequently reduced for 20- and 30-min etched ones. In a previous study, we have demonstrated in detail the dependency of SERS properties of SiNWs composites as a function of its etching time (Daoudi et al., 2020). Aforementioned, SiNWs with larger bundle size were formed at higher etching time significantly compromise the effective top surface area available for attaching

AgNPr and analyte molecules. In contrast, smaller sized bundles formed at lower etching time found to offer a better platform with enhanced top surface area while reduced NW length restricts migration of AgNPr to its bottom.

The effect of GO incorporation on SiNWs on SERS properties of composite substrates was investigated using R6G as an analyte molecule (Figure 3a). We found that AgNPr/GO/SiNWs composites demonstrated a much-pronounced SERS effect than the corresponding AgNPr/SiNWs composites. Raman peak intensity is also strongly dependent on the etching time of NWs. Unlike AgNPr/SiNWs, the best SERS efficiency has been observed for sensor fabricated with high etching time. In fact, AgNPr/GO/SiNWs prepared using 30-min etching time shows the optimum microstructure for getting the highest Raman enhancement. This could be accounted to the fact that larger sized bundles could facilitate better orientation of GO sheets, which in turn enhances the synergetic effects with plasmonic AgNPr as efficient hot spots.

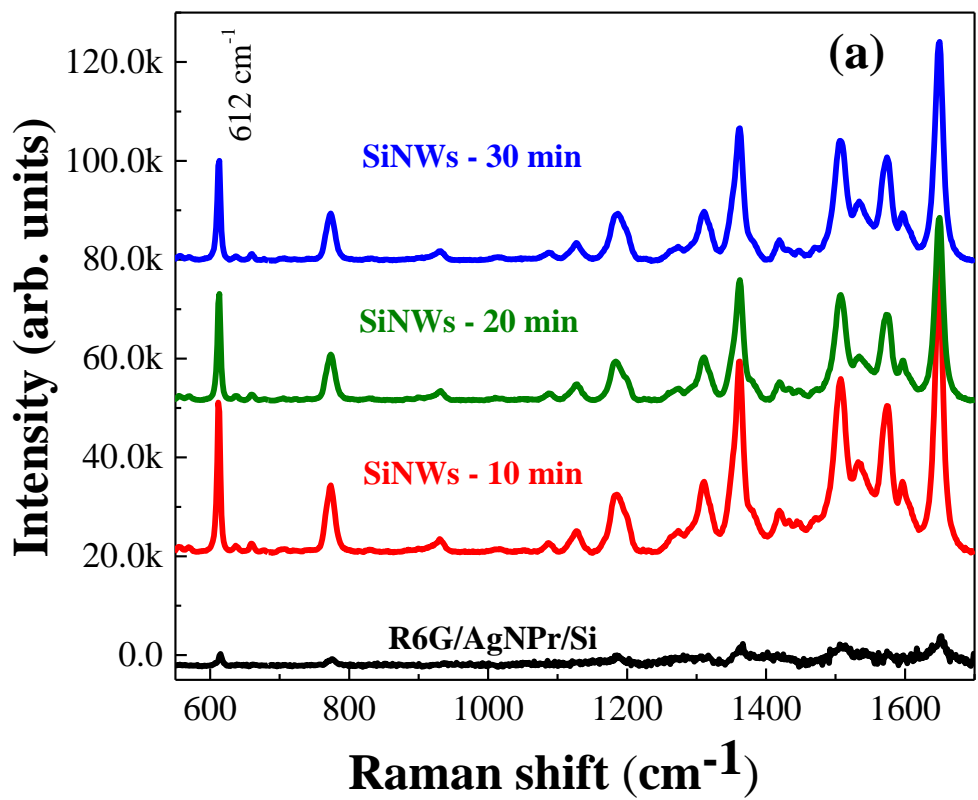
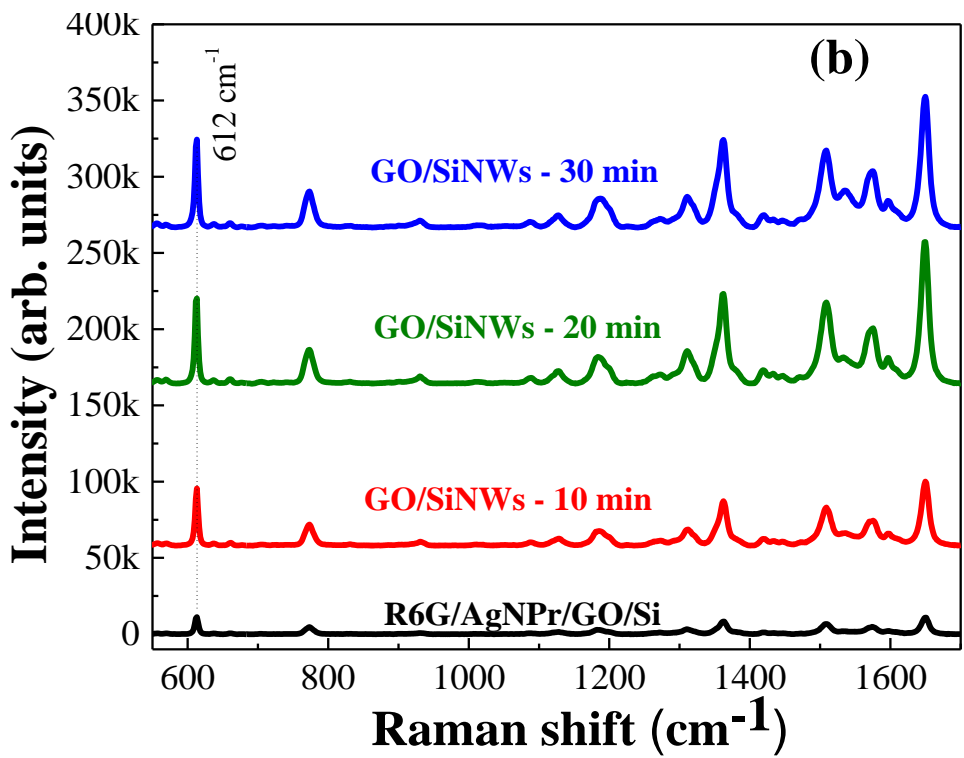

**Figure 3:** Raman spectra of R6G deposited substrates with varying etching time 10, 20, and 30 min (a) AgNPr/SiNWs and (b) AgNPr/GO/SiNWs.

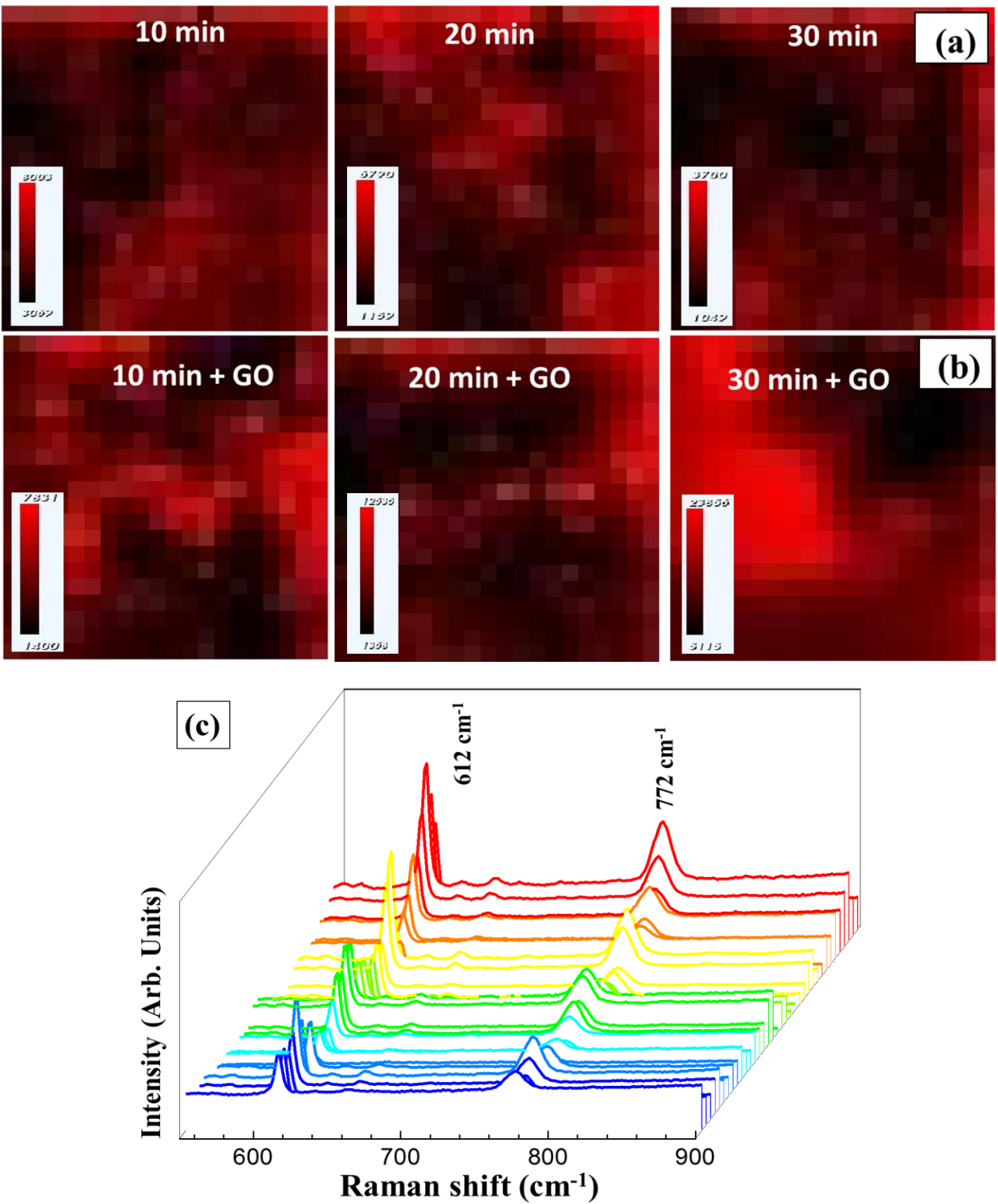

**Figure 4:** First and second row show the Raman maps of (a) R6G/AgNPr/SiNWs and (b) R6G/AgNPr/GO/SiNWs for various etching times (10 to 30 min). (c) 36 randomly chosen SERS spectra from SERS mapping results for the R6G/AgNPr/GO/SiNWs-10 min.

The distribution of Raman peak intensities of R6G within various substrates was investigated using Raman 2D mapping (Figure 4a & b). The red points and black points refer to maximum and minimum intensities respectively of Raman signal obtained on composite substrates. Since the morphology of Silicon nanowire can be compared to a hill and valley structure, the position of the laser pointer either on bundle top or bundle valley may affect the resultant intensity of the Raman signal. While the highest Raman signal enhancement was obtained on bundle top, the signal intensity was found to be lowered on bundle valleys. This could give rise to the observed inhomogeneity of Raman peaks (Figure 4c), which is reported to be associated with SiNWs nanostructure due to higher occurrence of AgNPr and R6G in bundle top and enhanced scattering loss at bundle valleys (Daoudi et al., 2020). After GO immobilization, the peaks intensity uniformity is improved while showing dependence on the etching factor (Fig 4b). On moving from lower to higher etched substrates, the area of black area (corresponding to minimum peak intensity) becomes more prominent which corresponds to bundle valleys. This effect compromised the SERS effect of AgNPr/SiNWs composites at higher etching times. However, one could observe that the area of red spots becomes bigger and denser for higher etched GO modified composites especially for 30-min etched samples.

As AgNPr/GO/SiNWs composites demonstrated the highest SERS response, these were selected for further sensitivity studies, which were conducted using different concentrations of R6G as shown in Figure 5. The peak centered at 612 cm$^{-1}$ were plotted with different concentrated R6G solutions (Figure 5a). It can be noted that AgNPr/GO/SiNWs composites could detect down to $10^{-12}$ M concentrations of R6G. The linear fitting curve (Figure 5b) also showed the corresponding linearity of the graph with an $R^2$ value of 0.98.

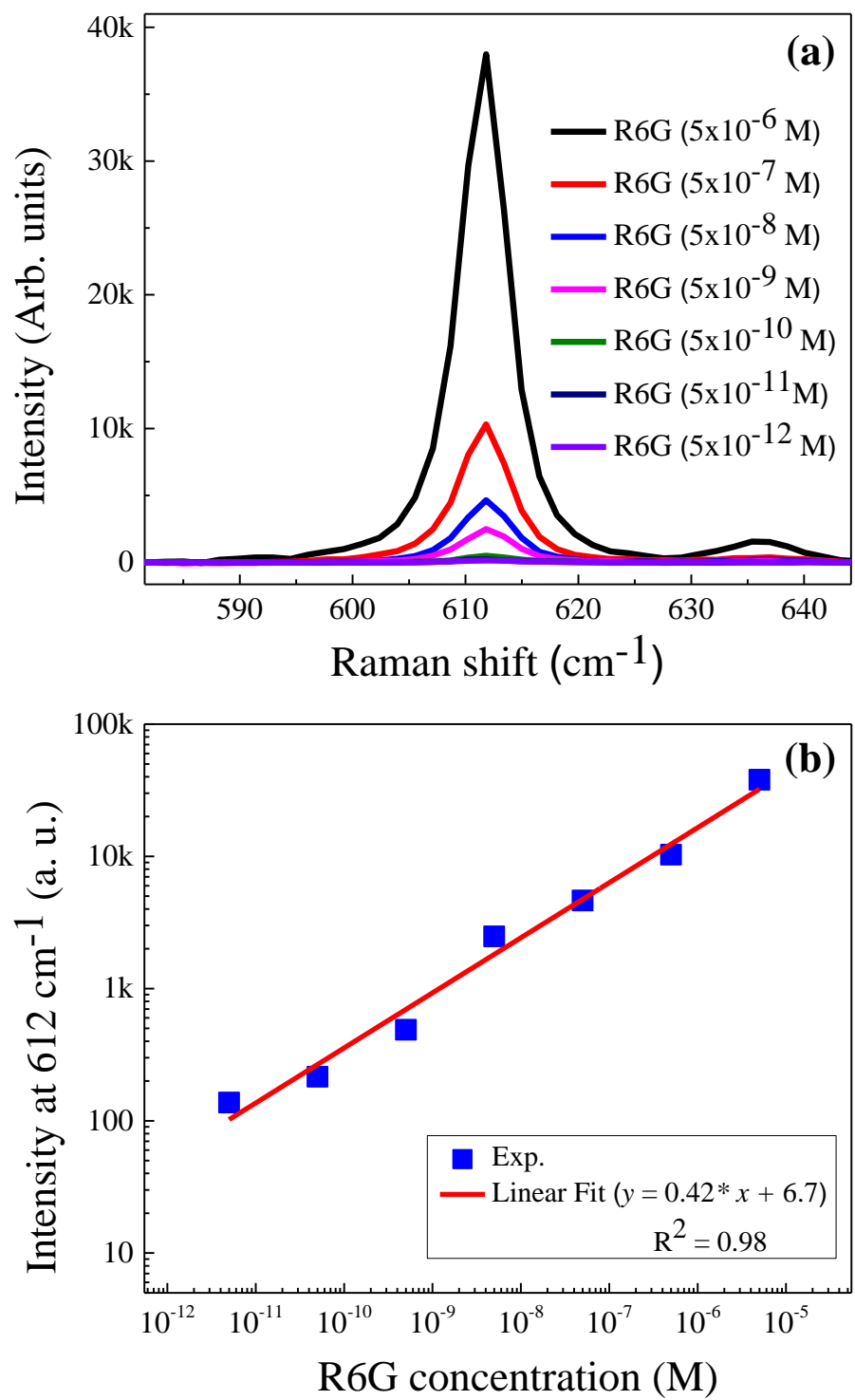

**Figure 5:** (a) Comparison of the Raman spectra centered at 612 cm$^{-1}$ for R6G/AgNPr/GO/SiNWs (30 min) with different concentration of R6G. (b) Raman intensity at 612 cm$^{-1}$ vs. R6G concentration in a log-log plane, along with linear fitting.

SERS enhancement factor (EF) is an important parameter that is generally calculated in order to characterize the proposed SERS sensor. The EF is usually described as the intensity ratio between SERS and normal Raman scattering for a given analyte normalized by the number of molecules probed. In the present study, the (EF) were calculated following the method used by Andrea *et al* (D'Andrea et al., 2016) for similar AgNPs/SiNWs nanocomposites. The EF is obtained by comparing the Raman and the fluorescence cross-sections, $\sigma_R$, and $\sigma_F$, respectively (Nie and Emory, 1997; Fazio et al., 2011). The signal intensities ratio per molecule and per watt power are related to the ratio between cross-sections as below:

$$\frac{I_{SERS}}{I_F} = EF \frac{I_{Raman}}{I_F} = EF \frac{\sigma_R}{\sigma_F} = EF \times 10^{-9} \qquad (1)$$

$$EF = 10^9 \frac{I_{SERS}}{I_F} \qquad (2)$$

where $I_{SERS}$ is the maximum intensity of the 612 cm$^{-1}$ mode for an excitation wavelength at 488 nm, and $I_F$ is the intensity of the fluorescence background intensities coming from the corresponding sample at the same wavelength. Figure 6 depicts the Raman spectra of the R6G/AgNPr/SiNWs (10 min) before (a) and after (b) removal of the background (BG) fluorescence. To ensure the consistency of our study, the values of $I_{SERS}$ and $I_F$ are obtained from the Raman spectra with its background fluorescence (Fig. 6 (a)). The presence of such background fluorescence was always considered as major handicap in resonance Raman spectroscopy (RRS) and SERS techniques (Matousek et al., 2002). It has been reported that in many cases the large intensity of the background can compromise the capability to get consistent and reproducible compositional information from some biological Raman spectra (Kourkoumelis et al., 2012). To follow the evolution of the SERS effects of our samples versus the substrate morphology and composition, we plotted all Raman spectra after BG fluorescence subtraction.

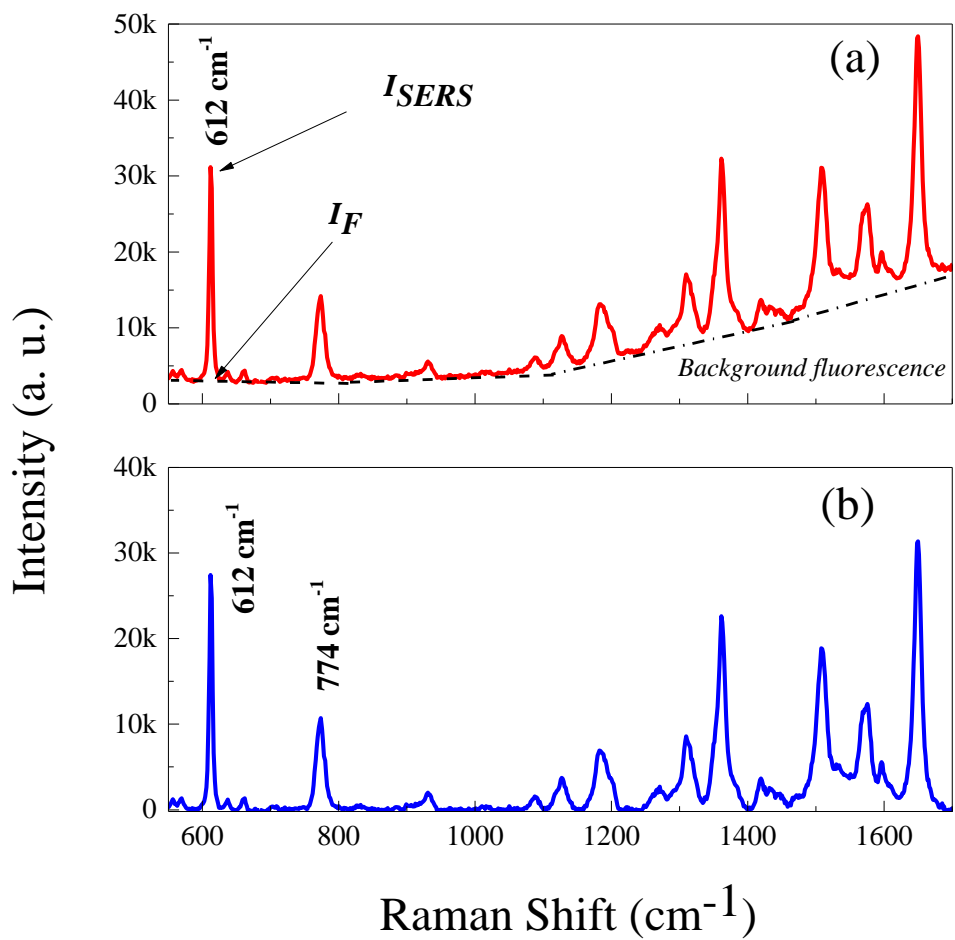

**Figure 6:** Raman spectra of the R6G/AgNPr/SiNWs (10 min) before (a) and after (b) the removal of the background fluorescence.

The values of the EF for SERS sensors based on AgNPr/SiNWs and AgNPr/GO/SiNWs obtained at various etching times are summarized in Table 1. In the case of AgNPr/SiNWs substrates, the highest SERS efficiency (3.2 x $10^{10}$) was obtained for 10 min and decreased for 20 and 30 min (Table 1). However, after modifying SiNWs with GO, the efficiency increases with the etching time. The largest efficiency value of 6.1 x $10^{10}$ has been obtained for R6G/AgNPr/GO/SiNW-30 min.

| Etch time | Enhancement Factor | |
|---|---|---|
| | AgNPr/SiNWs | AgNPr/GO/SiNWs |
| 0 min (bare Si) | $0.7 \times 10^{10}$ | $1.2 \times 10^{10}$ |
| 10 min | $3.2 \times 10^{10}$ | $3.9 \times 10^{10}$ |
| 20 min | $2.3 \times 10^{10}$ | $5.9 \times 10^{10}$ |
| 30 min | $2.2 \times 10^{10}$ | $6.1 \times 10^{10}$ |

**Table 1**: Enhancement factors of SiNWs based SERS composite substrates with and without GO for the detection of R6G molecule.

The improved SERS response for AgNPr/GO/SiNWs sensor could be related to the synergetic effects of AgNPr and GO. Studies suggested that while GO can provide good SERS efficiency by quenching background fluorescence (Li et al., 2000; Fan et al., 2014) the integration of a GO layer in the SERS device can protect the metal nanoparticles from oxidation (Zhang et al., 2017a; Naqvi et al., 2019). The high surface area of GO facilitates high absorption of the analyte molecules also. Reports also suggested that neighboring AgNPrs could be efficiently separated with GO layers and strongly enhance the intensities of hot spots via strong inter-coupling of LSP (Yi et al., 2016). Two main mechanisms could be responsible for SERS enhancement on graphene coated SiNWs. Firstly, charge transfer from the graphene to nanoparticle could lead to an enhanced electric field via LSP. The second mechanism is based on LSP confinement on graphene/SiNWs composite. The LSP confinement helps in electromagnetic waves focusing into smaller areas, which will further enhance the intensity of reflected electromagnetic field and hence SERS signal (Srichan et al., 2016).

### 3.4. SERS for the detection of methylene blue (MB) and Atrazine

SERS sensitivity studies using MB solution showed an increase in the intensity of characteristics peaks corresponding to MB with an increase in its concentration (Figure 7a). Characteristics peaks of MB at 449, 671, 951, 1154, 1622 cm$^{-1}$ were observed (Li et al., 2016a). In comparison to the peak at 521 cm$^{-1}$ corresponding to Si, the major peak at 1621 cm$^{-1}$ representing

MB was analyzed using $10^{-6}$ to $10^{-12}$ M concentrations. AgNPr/GO/SiNWs composites demonstrated an excellent low detection limit down to $10^{-12}$ M. Moreover, the linear relationship between the intensity of the Raman peak at 1621 cm$^{-1}$ and the concentration of MB solutions was elucidated (Figure 7b).

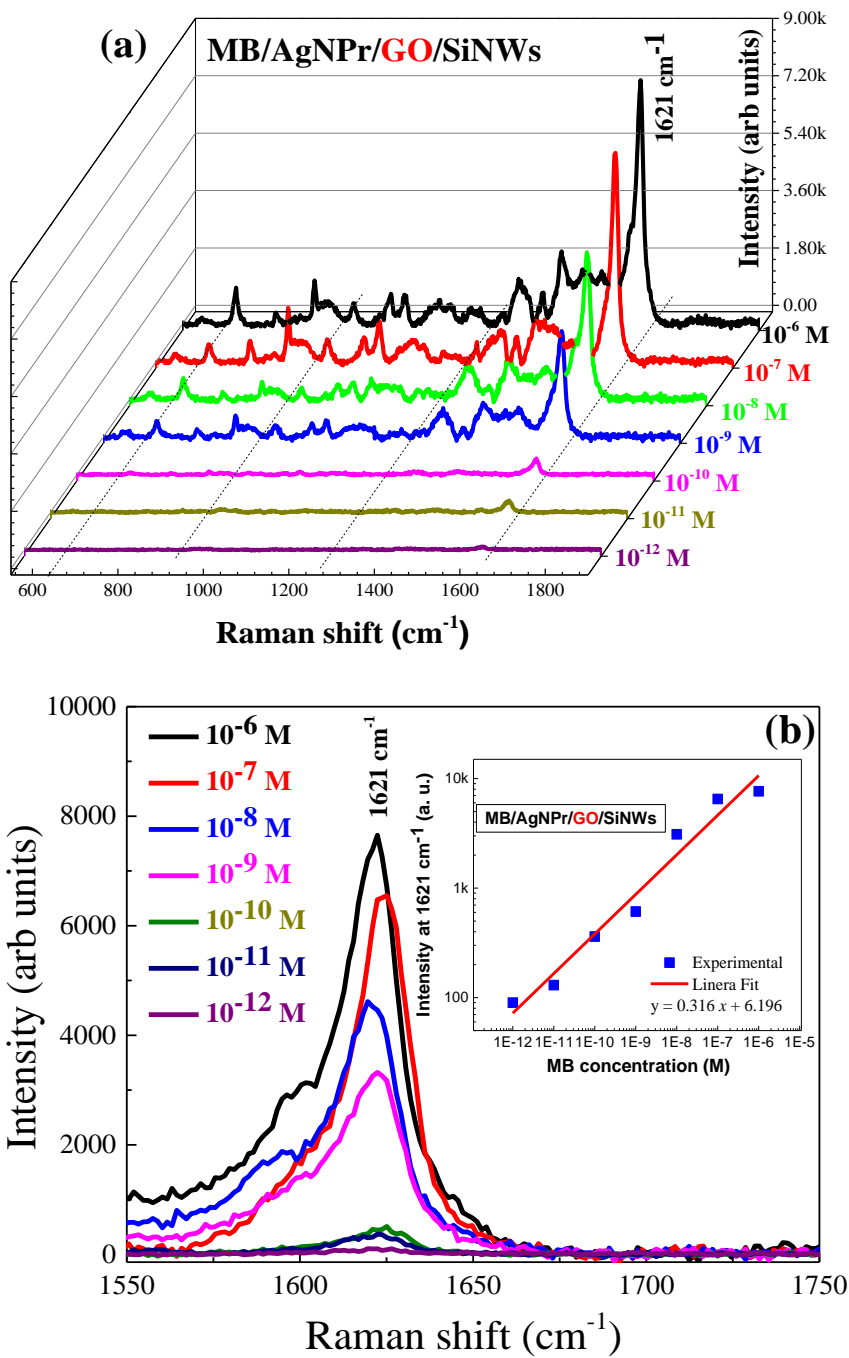

**Figure 7:** (a) Raman spectra collected for MB deposited on AgNPr/GO/SiNWs structures. (b) shows the evolution of the 1621 cm$^{-1}$ Raman mode vs. MB concentration. Inset of (b) shows the Raman intensity at 1621 cm$^{-1}$ vs. MB concentration in log-log plane, along with linear fitting.

Similar to the studies using MB, the sensitivity of the AgNPr/GO/SiNWs were further evaluated using atrazine also (Figure 8).

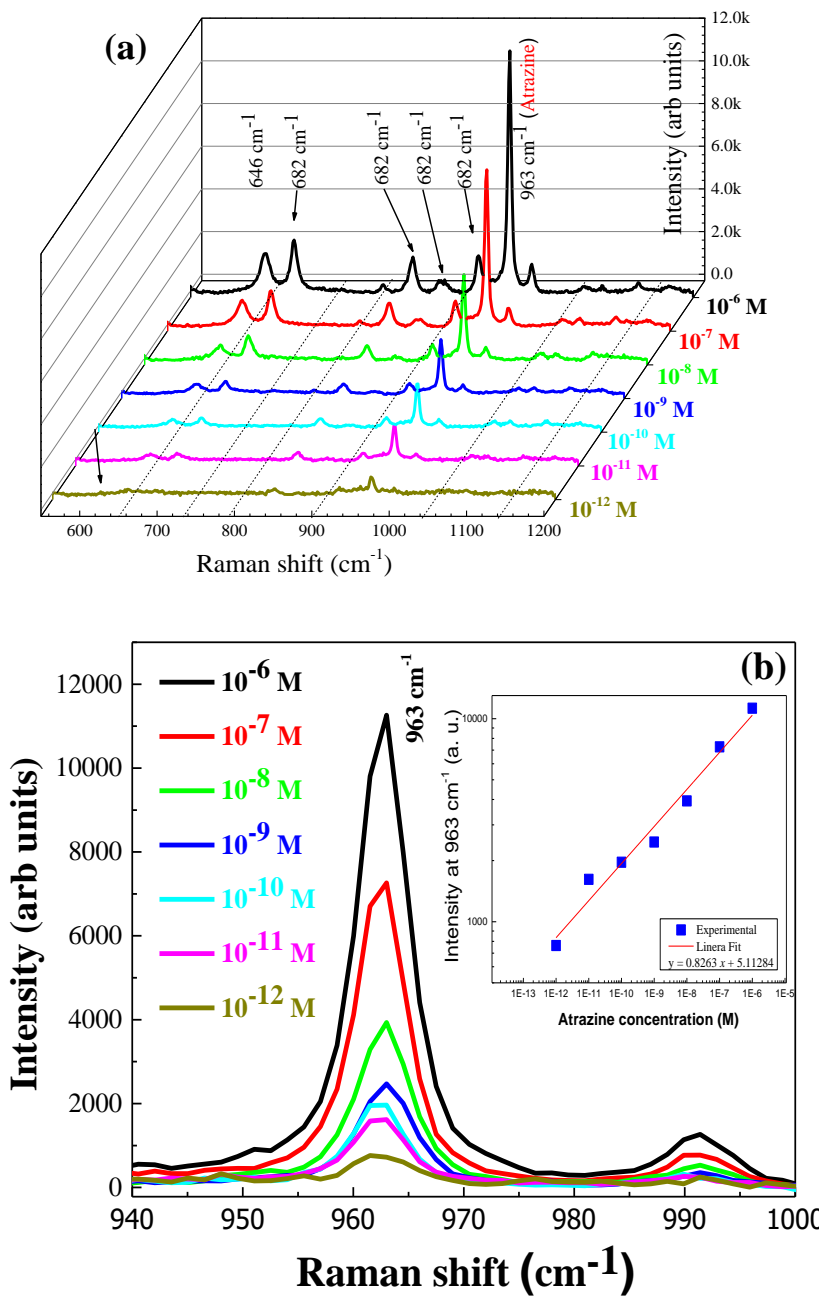

**Figure 8:** (a) Raman spectra recorded for atrazine deposited on AgNPr/GO/SiNWs structures. (b) shows the evolution of the 963 cm$^{-1}$ Raman mode vs. atrazine concentration. Inset of (b) shows the Raman intensity at 963 cm$^{-1}$ vs. Atrazine concentration in a log-log plane, along with linear fitting.

| Molecule/SERS substrate | Frequency (cm$^{-1}$) | Enhancement factor | Detection limit (M) | Reference |
|---|---|---|---|---|
| R6G/AgNPr/SiNWs | 612 | 3.2 x 10$^{10}$ | 5.0 x 10$^{-11}$ | This work |
| R6G/AgNPr/GO/SiNWs | 612 | 6.1 x 10$^{10}$ | 5.0 x 10$^{-12}$ | This work |
| AgNPs/Si | 612 | 10$^{10}$ | 10$^{-11}$ | (Tu et al., 2018) |
| R6G/AgNPs (PLD) / SiNWs | 612 | 3.0 x 10$^{8}$ | 1.0 x 10$^{-8}$ | (D'Andrea et al., 2016) |
| R6G / Gold nanoparticles | 1649 | 1.83 x 10$^{8}$ | - | (Yi et al., 2014) |
| MB/AgNPr/SiNWs | 1621 | - | 2.0 x 10$^{-11}$ | This work |
| MB/AgNPr/GO/SiNWs | 1621 | - | 2.0 x 10$^{-12}$ | This work |
| R6G / Gold film | 1365 | 1.0 x10$^{11}$ | - | (Xiao and Man, 2007) |
| MB/AgNPs/nodicaluminum oxide (AAO)template | - | - | 1.0 x 10$^{-8}$ | (Nuntawong et al., 2010) |
| MB/Au-Nanorod@SiO$_2$ | - | 3.0 x10$^{10}$ | - | (Seo et al., 2014) |
| MB/MoO$_3$/MoO$_2$ Nanosheets | 1630 | 1.4 x10$^{5}$ | 1.0 x 10$^{-9}$ | (Ren et al., 2020) |
| Atrazine/AgNPr/GO/SiNWs | 963 | - | 2.0 x 10$^{-12}$ | This work |

| Atrazine/AgNPs colloids | - | - | 5.0 x 10$^{-12}$ | (Rubira et al., 2014) |

**Table 2:** Comparison of our optimized values of the enhancement factors and detection limits with those reported in the literature for the detection of R6G, MB and Atrazine molecules using various SERS substrates.

Characteristics peaks such as 646, 682, 837, 871, 962, 990, 1250 cm$^{-1}$ were observed from the collected Raman spectra of atrazine solutions (Costa et al., 2011). The dependence of the intensity of peaks corresponding to atrazine on concentration was shown in Figure 8a. For 10$^{-6}$ M solutions, the peaks at 963 cm$^{-1}$ for atrazine showed high intensity with respect to peaks corresponding to Si and GO. Interestingly, the observed peaks are almost like those obtained for bare AgNPr/GO/SiNW (figure 2d) at the lowest detection levels (10$^{-12}$ M); especially demonstrated peaks for GO. AgNPr/GO/SiNWs showed atrazine detection levels up to 10$^{-12}$ M.

Table 2 demonstrated comparison of SERS efficiency and detection levels established with AgNPr/GO/SiNWs composites with the reported works for R6G, MB and atrazine. Compared to various SERS substrates, AgNPr/GO/SiNWs composites demonstrated high efficiency with excellent sensitivity for detecting both organic dyes and herbicide residues. In the current study, we attempted to demonstrate the relevance of etching time on microstructure of silicon nanowires composites and correlated to the corresponding SERS performance. Synergetic effects of plasmonic nanoparticles with graphene oxide combines with SiNWs proved to be an excellent strategy for developing simple and low-cost SERS substrates. The resultant NW composites found to have superior SERS response for detecting rhodamine 6G, methylene blue and atrazine residues.

**Conclusions**

SiNWs with modulated morphologies were successfully fabricated by varying etching time via silver assisted chemical etching technique. The micro-structural properties of AgNPr/GO/SiNWs composite matrix showed subsequent changes with increase in etching time from 10 to 30 min. Graphene oxide modification was found to improve the SERS properties of NW composites significantly. 30-min etched AgNPr/GO/SiNWs composites with optimum micro- topology exhibited highest SERS efficiency of 6.1 x 10$^{10}$. As-fabricated AgNPr/GO/SiNWs composites found be an excellent candidate for detection of various environmental pollutants such as organic dyes and herbicide residues up to $10^{-12}$ M.

**Acknowledgements**

The authors would like to acknowledge the financial support from the University of Sharjah and Sharjah Research Academy (Joint grant No. 802143072).